\documentclass[longnamesfirst,apjl]{emulateapj}
\usepackage{apjfonts,natbib,epsfig,bm,graphics,graphicx}
\newcommand{\sgra}{Sgr A$^*$}
\newcommand{\beq}{\begin{equation}}
\newcommand{\eeq}{\end{equation}}
\newcommand{\bea}{\begin{eqnarray}}
\newcommand{\eea}{\end{eqnarray}}
\begin{document}
\title{The Link between Warm Molecular Disks in Maser Nuclei
and Star Formation near the
Black Hole at the Galactic Center} 
\author{Milo\v s Milosavljevi\'c$^1$ and Abraham Loeb$^2$} 
\affil{ $^1$Theoretical Astrophysics, California Institute of Technology,
Pasadena, CA 91125, milos@tapir.caltech.edu; \\ $^2$Astronomy Department,
Harvard University, 60 Garden Street, Cambridge, MA 02138,
aloeb@cfa.harvard.edu }
\begin{abstract}
The discovery of hundreds of young, bright stars within a parsec from the
massive black hole Sagittarius A$^*$ at the center of the Galaxy presents a
challenge to star formation theories.  The requisite Roche densities for
the gravitational collapse of gas clouds are most naturally achieved in
accretion disks.  The water maser sources in Keplerian rotation in the
nuclei of NGC4258, NGC1068, and the Circinus galaxy indicate the presence
of warm, extended, molecular accretion disks around black holes similar in
mass to \sgra.  Here we argue that the current conditions in the maser
nuclei, and those near the Galactic center, represent two consecutive,
recurrent phases in the life cycle of the nucleus of a typical gas-rich
spiral bulge.  The warm molecular disks that give rise to the observed
maser emission fragment into stellar-size objects.  The stellar masses,
their orbital geometry, and the total number of stars thus formed are
consistent with the values identified at the Galactic center.  The stars
tend to form in compact groups resembling the IRS 13 complex that
dominates the stellar light in the neighborhood of \sgra.
\end{abstract}


\section{Introduction}

The dynamical sphere of influence of the massive black hole (MBH) at the
Galactic center, measuring 1 pc in radius, is host to a large number of
luminous, massive, and hence young stars (e.g.,~\citealt{Krabbe:95,Genzel:00}).
The closest star to the MBH with detailed proper motion data, S0-2, has an
apocenter distance of only $\sim0.01\textrm{ pc}$ and an eccentricity of
$\sim 0.88$ (e.g.,~\citealt{Schoedel:02}).  It is a B0-O8 dwarf with a
mass of $10-15M_\odot$ and an age of $\lesssim 10\textrm{ Myr}$
\citep{Ghez:03a}.  A number of other stars at radial distances $\lesssim
0.03\textrm{ pc}$ exhibit properties similar to those of S0-2
(e.g.,~\citealt{Ghez:03b}).  At a projected distance $\gtrsim 0.1\textrm{
pc}$ from the MBH, a cluster containing seven blue supergiants, known as the
IRS 13 complex, dominates the ionizing luminosity
(e.g.,~\citealt{Najarro:97}).  At somewhat larger distances of
$0.1-0.5\textrm{ pc}$, as many as $\sim 40$ Wolf-Rayet stars have been
observed \citep{Genzel:03}.  Since the zero-age main sequence (ZAMS) mass
of the Wolf-Rayet stars is $30-100M_\odot$, this implies that the Galactic
center had been host to a recent starburst with a total initial stellar
mass of greater than $10^3M_\odot$.

Various dynamical mechanisms by which young stars could have migrated from
larger radii into a fraction of the central parsec have been considered
\citep{Gerhard:01,Gould:03,KimS:03}. 
None are successful at explaining the observed
concentration of young stars, unless an additional, hitherto undetected
dynamical component, such as an intermediate-mass black hole, is invoked
\citep{Hansen:03}.  This conclusion, therefore, suggests that the stars
have probably formed in-situ.  Molecular cloud densities of
$10^{14}\textrm{ cm}^{-3}$ at the radius corresponding to the apocenter
distance of S0-2, and those of 
$10^{9}-10^{10}\textrm{ cm}^{-3}$ at the orbital
radii of the Wolf-Rayet stars, are required for gravitational collapse to
occur in the presence of the Sagittarius A$^*$ 
tide \citep{Morris:93}.  The densities
of molecular cloud fragments found in the central 2 pc of the Galaxy,
however, do not exceed $\sim10^6\textrm{ cm}^{-3}$
(e.g.,~\citealt{Jackson:93}).

The densities necessary for gravitational collapse in the tidal field of
the MBH could easily be sustained in accretion disks.  The young stars may
be connected to a past accretion disk at the Galactic center since their
orbits appear to lie within one or more distinct planes
\citep{Levin:03,Genzel:03}.  Multiple stellar disks have been observed in
the nuclei of other galaxies at the resolution limited radii of
$\sim20\textrm{ pc}$ (e.g.,~\citealt{Pizzella:02}).

The proposal that stars form in the outer parts of accretion disks of
active galactic nuclei (AGNs) is not new \citep{Kolykhalov:80}. The disks
are susceptible to self-gravitating instability when the
\citet{Toomre:64} parameter $Q=c_s\Omega/\pi G\Sigma$ is less than unity,
where $c_s$ is the sound speed, $\Omega=(GM_{\rm bh}/r^3)^{1/2}$ is the
angular velocity (assuming a Keplerian potential), $M_{\rm bh}$ is the mass
of the MBH, and $\Sigma$ is the surface density of gas.  For a range of
disk parameters, the disk fragments into stellar-size clumps.
\citet{Shlosman:89} discussed star formation in cold, molecular
disks of AGN.  Grains in the surface layer of a disk absorb the incident
optical-to-soft X-ray flux from the AGN and reradiate it in the infrared.
The disk bulk temperature remains low; e.g., if the $4\times10^6M_\odot$
black hole at the Galactic center were accreting at the Eddington rate
during a past accretion episode, it would have heated the disk to
temperatures $\sim 36 r_3^{-1/2}\textrm{ K}$, where $r=0.3 r_3 \textrm{
pc}$ is the distance from the MBH.  Such a cold disk is self-gravitating at
surface densities $\gtrsim 5\textrm{ g cm}^{-2}$.  The apparent
inevitability of fragmentation implies a crisis for the extended disk
models (\citealt{Goodman:03}).

While the construction of extended disk models may be challenging, nature
offers evidence that these disks indeed exist.  Maser emission in the
rotational level transition $6_{16}\rightarrow5_{23}$ of water has been
detected at 22 GHz in the nuclei of several Seyfert II and LINER galaxies.
Keplerian rotation patterns and disk-like geometries of the maser sources
at distances $\gtrsim0.1\textrm{ pc}$ from the MBH have been detected in
(at least) three cases: NGC4258 \citep{Miyoshi:95,Greenhill:95}, NGC1068
\citep{Gallimore:96}, and the Circinus galaxy
\citep{Greenhill:03a,Greenhill:03b}.  This has provided an accurate mass
determination for the central black holes, as well as the distances to some
host galaxies (e.g., \citealt{Herrnstein:99}).

In Table \ref{tab:params}, we summarize the inferred values of the black
hole mass $M_{\rm bh}$, the radii of the masing sources from the MBH
$r_{\rm disk}$, and the stellar velocity dispersions $\sigma_{\rm stellar}$
of the host bulges.  Note that in all cases the masers are located within
the spheres of dynamical influence of the MBH $r_{\rm disk}<GM_{\rm
bh}/\sigma_{\rm stellar}^2$, which reflects the selection of these sources
for nearly-Keplerian rotation patterns.

The purpose of this Letter is to highlight a connection between the
warm molecular disks in AGNs and the stellar populations at the Galactic
Center.  In \S~\ref{sec:disk}, we review the thermodynamical properties of a
warm molecular disk by assuming that the conditions in this disk are uniformly
conducive to the production of water maser emission.  In
\S~\ref{sec:stars}, we study the fragmentation of the disk and the
formation of stars.

\begin{deluxetable}{cccc}
\tablecolumns{4}
\tablehead{
  \colhead{Galaxy} & 
  \colhead{$M_{\rm bh}$ ($10^7 M_\odot$)} & 
  \colhead{$r_{\rm disk}$ (pc)} & 
  \colhead{$\sigma_{\rm stellar}$ (km s$^{-1}$)} 
}
\startdata
NGC4258   & 3.9\tablenotemark{a}  & $0.16-0.28$\tablenotemark{a}  & 167\tablenotemark{f}               \\
NGC1068   & 1.0\tablenotemark{b}  & $>0.6$\tablenotemark{c}       & 177\tablenotemark{f}      \\
Circinus  & 0.17\tablenotemark{d} & $0.11-0.4$\tablenotemark{d} & 167\tablenotemark{f}                        \\
Milky Way & 0.4\tablenotemark{e} &  & $75-125$                   \\
\enddata
\tablenotetext{a}{\citet{Herrnstein:99}}
\tablenotetext{b}{\citet{Greenhill:96}}
\tablenotetext{c}{\citet{Gallimore:01}}
\tablenotetext{d}{\citet{Greenhill:03b}}
\tablenotetext{e}{\citet{Ghez:03b}}
\tablenotetext{f}{\citet{Hypercat}}
\label{tab:params}
\end{deluxetable}

\section{Conditions in the Disk from the Requirements for Maser Action}
\label{sec:disk}

An inverted population of the maser levels, which is a precondition for maser
action, requires a minimum gas temperature of $400\textrm{ K}$
(e.g.,~\citealt{Maloney:02,Watson:02}).  Collisional pumping of the
levels requires that the density of the masing gas be at least
$10^7\textrm{ cm}^{-3}$.  Above $10^{10}\textrm{ cm}^{-3}$, the levels are
thermalized, and the maser action is quenched.  We assume that the
conditions for the operation of masers are homogeneously represented in the
disk, and we adopt the fiducial temperature $T_{\rm gas}=400\textrm{ K}$ and
density $n_{\rm gas}=10^9\textrm{ cm}^{-3}$ at a distance $r_{\rm disk}\sim
0.3\textrm{ pc}$ from a central black hole of mass $M_{\rm bh}\sim
10^7M_\odot$.  We defer discussing the origin of the unexpectedly
high temperature of the molecular gas to \S~\ref{sec:discussion}.

The assumption that the accretion disk 
has on average the same parameters as the
masing regions is not essential; i.e., the masing regions could be 
localized condensations in the disk (\S~\ref{sec:discussion}). 
The star formation mechanism presented in \S~\ref{sec:stars}, however, 
is valid even at average temperatures of less than $400\textrm{ K}$.

The vertical scale height of a disk in pressure equilibrium is
$H=c_s/\Omega=\Sigma/\mu m_{{\rm H}_2} n_{\rm gas}$, where the 
isothermal sound speed
of gas with a mean molecular weight $\mu$ in units of the hydrogen molecular
mass is related to the temperature via ${c_s}^2=kT_{\rm gas}/\mu m_{{\rm
H}_2}$.  
The vertical column density of nuclei amounts to 
$\Sigma/m_{\rm H}\approx
8.5\times10^{24} \mu^{-1/2}n_9 T_4^{1/2} r_3^{3/2} M_7^{-1/2}\textrm{
cm}^{-2}$, where $n_{\rm gas}=n_9\times10^9\textrm{ cm}^{-3}$, $T_{\rm gas}=400
T_4\textrm{ K}$ and $M_{\rm bh}=M_7\times10^7M_\odot$.

Note that very similar column densities of $10^{25}\textrm{ cm}^{-2}$ and
$4\times10^{24}\textrm{ cm}^{-2}$ have been inferred from hard X-ray
absorption measures in the nuclei of NGC1068 \citep{Matt:03}
and the Circinus galaxy \citep{Matt:99}, respectively.  
This coincidence lends support to
our characterization of the conditions in the disk.  Detailed comparison,
however, is not possible because the inclination and the corresponding
absorbing column densities depend sensitively on the unknown degree of warping
in the observed disks.

\section{Formation of Stars in a Warm Molecular Disk}
\label{sec:stars}

Since $H/r\sim 0.003 (T_4 r_3/\mu M_7)^{1/2}$, the disk is geometrically
thin, and its gravitational stability can be determined by evaluating the
Toomre parameter $Q=2.6n_9^{-1}r_3^{-3}M_7$.
We find that $Q$ drops below unity outside
the critical radius of $\sim0.42 (M_7/n_9)^{1/3} \textrm{ pc}$.  The critical
radii coincide, to within the uncertainty in $n_{\rm gas}$, with the radii
of the observed maser disks.
 
Fragmentation of a marginally unstable disk produces gravitationally bound
objects of the Jeans mass $M_{\rm Jeans}=c_s^4/G^2\Sigma
\sim3\mu^{-5/2} n_9^{-1}T_4^{3/2} r_3^{-3/2} M_7^{1/2} M_\odot$.
The Jeans length is approximately equal to the disk
thickness, which is in turn equal to the Hill (Roche) radius $R_{\rm
H}=(m_*/3M_{\rm bh})^{1/3}r$ of a newly formed self-gravitating
object of mass $m_*$ (hereafter a ``protostar'').

The total disk mass $\pi r_{\rm disk}^2\Sigma$ within the
radius of the masers is
$M_{\rm disk}\sim 1.3\times 10^4 \mu^{1/2}
n_9 T_4^{1/2} r_3^{7/2} M_7^{-1/2} M_\odot$.
This mass is comparable to the conservative estimates of the
mass of the circumnuclear disk (CND) at the Galactic center
(\citealt{Mezger:96} and references therein).  

The maximum number of separate protostars that can be produced by the Jeans
instability is then $N_{\rm max}\sim M_{\rm disk}/M_{\rm Jeans}\sim4300 f
\mu^3 {n_9}^2 {r_3}^5/T_4 M_7$, where $f$ is the fraction of the disk mass
that undergoes fragmentation.  We expect that the latter fraction is well
below unity, e.g., $f\sim 0.1\%-1\%$.  The density wakes associated with the
Lindblad resonances of the protostars that formed first stimulate further
fragmentation \citep{Armitage:99,Lufkin:03}.

Gas interior to the Hill annulus $|\Delta r|\lesssim R_{\rm H}$ in the disk
is delivered by differential rotation into the protostar's Roche lobe and
can accrete onto the star.  The final mass of the protostar can thus be
much larger than the Jeans mass.  Equating the total mass in the annulus
$2\pi r R_{\rm H}\Sigma\propto m_*^{1/3}$ to the mass of the protostar
$m_*$ and solving for the latter yield the maximum ``isolation mass''
\citep{Lissauer:87} $M_{\rm iso}\sim(4/3\pi)M_{\rm disk}^{3/2}/M_{\rm
bh}^{1/2}$ to which the star can grow by accreting the disk material.  In
gaseous protoplanetary disks, the growth of a planet terminates when a gap
of width $\gtrsim R_{\rm H}$ opens in the disk.  In AGN disks, however,
$M_{\rm iso}$ is a large fraction of $M_{\rm disk}$ \citep{Goodman:04}.

If multiple Jeans patches of the disk collapse simultaneously, their Hill
annuli grow until they overlap.  The accretion is terminated either by
reaching the isolation mass or by the global depletion of gas from the
accretion onto multiple protostars.

Before the protostellar mass has
reached the terminal value, differential rotation in the disk delivers disk
material into the Hill sphere of the protostar at the rate $\dot m_* \sim
R_{\rm H}^2 \Sigma \Omega$, or 
\bea
\label{eq:protostaraccret}
\dot m_* &\sim& (10^{-4} M_\odot \textrm{ yr}^{-1}) \mu^{1/2}
\left(\frac{m_*}{3M_\odot}\right)^{2/3} \left(\frac{n_{\rm
gas}}{10^9\textrm{ cm}^{-3}}\right) \nonumber\\ & &\times
\left(\frac{T_{\rm gas}}{400\textrm{ K}}\right)^{1/2} \left(\frac{r_{\rm
disk}}{0.3\textrm{ pc}}\right)^{2} \left(\frac{M_{\rm
bh}}{10^7M_\odot}\right)^{-2/3} .  
\eea 
If the protostar could accept mass at this rate, 
its growth time would be given
by $m_*/\dot m_*\sim 2.5\times 10^4\mu^{-1/2}m_3^{1/3} M_7^{2/3}r_3^{-2}
n_9^{-1} T_4^{-1/2}\textrm{ yr}$. 

The gas arriving into the Hill sphere undergoes a shock near the
L1 and L2 Lagrangian points and forms two streams around the protostar
\citep{Lubow:99,Bate:03}.  
The streams circularize at a fraction of $\sim 50\%$ 
of $R_{\rm H}$ from the protostar.  The protostar is 
therefore fed from an accretion disk of its own.  We have made an implicit
assumption that the entire mass entering the Hill sphere on horseshoe orbits 
is inelastically captured and remains inside the sphere.

Gas in the protostellar disk must be able to dispense its angular momentum
if it is to contract onto a central protostellar core.  The radial extent
of the protostellar disk is large, $\log(R_{\rm H}/2R_*)\gtrsim 4$, where
$R_*$ is the radius of a star of mass $m_*$ on the ZAMS.  In a purely
gaseous disk, angular momentum extraction could be achieved via radial
transport (``$\alpha$-viscosity''; \citealt{Shakura:73}) driven by the
magnetorotational instability (MRI; \citealt{Balbus:98}), via the magnetic
breaking (MB) mechanism (e.g.,~\citealt{Mouschovias:79}), or via the
centrifugal launching of winds from the disk surface \citep{Blandford:82}.

MRI and MB are expected to operate in the outermost region of the
protostellar disk if the concentration of free electrons in the gas is
sufficiently large to provide for an inertial coupling with a magnetic
field.  This concentration depends on the detailed chemistry that we do not
attempt to analyze here.  We are thus not in the position to decide whether
the outer edge of the protostellar disk is magnetically ``dead'' or
``active.''  If it is dead, the material captured inside the Hill sphere
accumulates in a violently unstable ring around the protostar.  If it is
active, it is, in principlem possible that a magnetically mediated angular
momentum extraction maintains gravitational stability in the protostellar
disk.  For example, \citet{Papaloizou:03} found
evidence of MB while simulating the accretion onto planetary embryos
embedded within a protostellar disk with ideal magneto-hydrodynamics (IMHD).
We proceed to check for gravitational stability assuming, optimistically,
that IMHD is realized in the protostellar disk.
  
If MRI is the dominant angular momentum extraction channel, and if the
accretion rate is given by equation (\ref{eq:protostaraccret}), the disk
parameters can be evaluated in the standard fashion
(e.g.,~\citealt{Frank:02}).  In the optically thick, adiabatic limit, we
have $\alpha=0.01\alpha_{-2}$ with $\alpha_{-2}\sim 0.1-1$
(e.g.,~\citealt{Sano:03}).  The Rosseland mean opacity of molecular gas
above $100\textrm{ K}$ and below the opacity gap at $\sim1000\textrm{ K}$
is in the range $\kappa\sim 2-10\textrm{ cm}^2\textrm{ g}^{-1}$
\citep{Pollack:94,Semenov:03}.  Assuming that the outer radius of the disk
is $\sim R_{\rm H}$, the Toomre parameter there equals $Q_*\sim 0.0075
\mu^{-1.4} \alpha_{-2}^{0.7} \kappa^{0.3} M_7^{0.72} r_3^{-2.2}
m_3^{-0.27} n_9^{-0.4} T_4^{-0.2}$.  (All other
parameters pertain to the parent, AGN disk medium.)  The outer edges of an
$\alpha$-protostellar disk are thus gravitationally unstable in their own
right.

If MB is the dominant angular momentum extraction channel, the angular
momentum is removed in a vertical Alfv\'en crossing time of the disk,
$t_{\rm mb}\sim\phi^{-1}\rho_{\rm gas}^{-1/2}$, where $\phi\sim B_{\rm
unif}/\Sigma$ is the specific magnetic flux threading the gas in the donor
AGN disk, $\rho_{\rm gas}$ is the ambient gas density (here assumed to
equal the density of gas in the donor disk), and $B_{\rm unif}$ is the net
magnetic field threading the disk.  If $B_{\rm unif}$ is a fraction $\xi$
of the equipartition field $(4\pi\rho_{\rm gas})^{1/2}c_s$, then $\phi=2\xi
\Omega(\pi/\rho_{\rm gas})^{1/2}$.  From mass and flux conservation in the
flux-freezing approximation, we get that the column density in the
protostellar disk $\Sigma_*$ is related to that in the AGN disk via
$\Sigma_*=\Sigma/\xi$. Thus, assuming a uniform gas temperature, the Toomre
parameters of the two disks are related via $Q_*=\xi Q$.  Since $Q\sim1$ and
$\xi\ll 1$, we expect $Q_*\ll 1$. Again, the protostellar disk is
gravitationally unstable.

We have found that the protostellar disk is susceptible to fragmentation in
its own right and separates into multiple clumps. This process has recently
been identified in simulations
\citep{Boss:02,Nakamura:03,Machida:03}.  While the Hill sphere
continues to receive gas at the rate given by equation 
(\ref{eq:protostaraccret}), this gas
does not accrete to a single protostar but to a group ($N_{\rm
group}\geq2$) of protostars sharing the same Hill sphere.  If $N_{\rm
group}\ga 3$, the spatial extent of the group increases because of the strong
two-body encounters (``dynamical evaporation'') beyond the original Hill
sphere.  The evaporation takes place on a timescale of $\sim
\Omega^{-1}N_{\rm group}$.  The final group does not share the same Hill
sphere even if the initial group did, and it is tidally stretched.

Evidence that stars form in compact groups can be found at the Galactic
center.  The IRS 13 complex is located at the projected distance of $\sim
3''\approx0.12\textrm{ pc}$ from \sgra and has an apparent diameter
of less than 
$0.04\textrm{ pc}$. This compact stellar cluster contains seven blue
supergiants. Recently, additional candidate members of this or
a similar group have been
discovered \citep{Eckart:03}.  Assuming an inclination of IRS 13 relative
to \sgra of $45^\circ$ and an initial Hill radius of $\sim 0.01$ pc, the
spatial extent of IRS 13 would correspond to a Hill sphere associated with
a total mass of $\sim 2500M_\odot$.  This is a fraction of the mass of the
present CND and is compatible with the isolation mass limit for a $3\times
10^4 M_\odot$ disk that would form if the CND were to start accreting
toward the MBH.

The stars forming in the disk possess initially quasi-circular orbits, yet
several of the most bound stars, including S0-2, exhibit high
eccentricities $e\gtrsim 0.8$.  While the disk torques are not effective at
generating such high eccentricities, nearby encounters between stars could,
in principle, scatter stars onto radial orbits.  An encounter between stars
initially belonging to disjoint Hill annuli induces a velocity change $\Delta
v\sim (G m/R_{\rm H})^{1/2}\sim (m/M_{\rm bh})^{1/3} r\Omega$, which is
much too low
to deflect the star by of order its own velocity, $\sim r\Omega$,
required for $\Delta e\sim 1$.  Significantly larger velocity kicks result
from three- or four-body gravitational slingshot interactions between
{\it binary} stars sharing the same Hill sphere (massive stars in the field
generically form in compact binaries; e.g.,~\citealt{Vanbeveren:98}).
Indeed, the circular velocity at S0-2's apoapse, $\sim 1300\textrm{ km
s}^{-1}$, is comparable to the maximum velocity that a $15M_\odot$ star can
be deflected by, while avoiding direct collision with a similar star.

\section{Discussion}
\label{sec:discussion}

So far, we have avoided explaining how the high temperature
($\sim400\textrm{ K}$) can be maintained in the maser disk.  Dissipation of
turbulence within the disk is not effective in this regard
\citep{Neufeld:95,Desch:98}.  To demonstrate this, we search for the
(unrealistic) value of $\alpha$ that would be necessary to power the masing
disk from within. The vertical optical depth equals $\tau= \kappa
\Sigma\sim10\mu^{1/2}\kappa n_9 T_4^{1/4} M_7^{-1/2}r_3^{3/2}$.  
The accretion time
$t_{\rm acc}=\pi r^2\Sigma/\dot M$ is larger by $\time10^3-10^5$
than the dynamical time. Thus, the accretion rate $\dot M=3\pi\alpha
c_s^2\Sigma/\Omega$ may be very different at different radii
\citep{Gammie:99} and need not be related to the bolometric luminosity of
the AGN.  Instead, we assume that $n_{\rm gas}$ and $T_{\rm gas}$ are
independent variables and obtain $\alpha\sim750\kappa^{-1}n_9^{-1}
{T_4}^2 M_7^{1/2} r_3^{-3/2}$.  This requirement is orders of magnitude
larger than the value typically found in numerical simulations.  Therefore,
the standard $\alpha$-disks are not sufficiently warm to form masers.

Next, we check whether or not the protostellar radiative output could heat the
disk.  Part of the residual disk gas is accessible to the ionizing
radiation from the new stars.  Stars with ZAMS masses $\gtrsim8M_\odot$
ionize the ambient hydrogen within $\sim10^5\textrm{ yr}$ from their
initial collapse (e.g.,~\citealt{Yorke:86}), and create \ion{H}{2} regions
within their parent disk.  The characteristic region size is given by the
Str\"omgren radius $R_{\rm S}=(3{\cal F}/4\pi n_{\rm H}^2\alpha_B)^{1/3}$,
where ${\cal F}$ is the number of ionizing photons per unit time emitted by
the star and $\alpha_B\approx(2.6-4.5)\times10^{-13}\textrm{ cm}^3\textrm{
s}^{-1}$ is the case-B recombination coefficient of hydrogen 
\citep{Osterbrock:89}.  The ionizing photon number flux for a star with
an effective temperature $T_*$ equals ${\cal F} \approx 16\pi^2kT_*
R_*^2\nu_0^2 c^{-2} h^{-1} e^{-h\nu_0/kT_*}$, where $h\nu_0=13.6\textrm{
eV}$.  Taking the stellar luminosity $L_*\approx(15,000L_\odot){m_8}^3$,
where $m_8=m_*/8M_\odot$ and stellar radius $R_*\approx(3.2 R_\odot)
{m_8}^{0.56}$, we get $R_{\rm S}\approx
4\times10^{14}\ \sqrt{m_8}e^{-1.5/\sqrt{m_8}} n_9^{-2/3}\textrm{ cm}$.  

The \ion{H}{2} region is much smaller than the disk radius even if $\sim 99\%$
of the disk gas has been depleted.  The resulting \ion{H}{2} distribution
has a ``blister'' geometry typically found in the Orion Nebula and in other
star-forming regions (SFRs).  In SFRs, maser sources are found in the
shocked, dense shell of the blister.  The isotropic luminosities of
the SFR masers, $10^{-3}$ to $1L_\odot$, are much smaller than those of the
circumnuclear masers, $10$ to $10^3L_\odot$. This may be the result of a
longer gain path along coherent velocity patches in the AGN disk (where
the gas flow is organized) rather than a larger pumping luminosity 
(e.g.,~\citealt{Deguchi:89}).

The stars could also heat the disk gas mechanically,
by driving spiral density waves that in some circumstances 
lead to spiral shocks.
Such coherent, spiral shocks in the disk have
been proposed as the sites of maser emission \citep{Maoz:98}.  
This explanation, however, depends on many details of the model 
that we do not attempt to investigate here. 
The most promising mechanism for heating the disk gas may still be
illumination by the X-ray flux of the AGN \citep{Neufeld:94}. 
The X-rays heat a surface layer in the disk while
allowing the midplane to remain cooler.  If the average
midplane temperature of the disk were, say, $\sim 5$ times lower then the
Jeans mass would have been, $M_{\rm Jeans}\sim 0.27M_\odot$.  The subsequent
evolution described in \S~\ref{sec:stars}, including the rapid accretion of
disk gas, the subfragmentation into multiple stars sharing the same Hill
sphere, and the global depletion of disk gas, would occur in the same
fashion.

\acknowledgements

We thank A.~Barth, R.~Blandford,
P.~Goldreich, J.~Moran, S.~Phinney, F.~Rasio, 
R.~Sari and R.~Sunyaev for
inspiring discussions and W.~Watson for valuable comments.  
The research of M.~M. was supported at Caltech by a
postdoctoral fellowship from the Sherman Fairchild Foundation.  This work
was also supported in part by NASA grant NAG 5-13292, and by NSF grants
AST 00-71019 and AST 02-04514 to A.~L.

\end{document}